\documentclass[aps,prx,a4paper,twocolumn,superscriptaddress,longbibliography,floatfix]{revtex4-2}

\usepackage{amssymb,amsmath,amsfonts}
\usepackage{graphicx}
\usepackage[dvipsnames]{xcolor}
\usepackage[bookmarksopen,colorlinks,linkcolor=blue,citecolor=olive,urlcolor=red]{hyperref}
\usepackage{bm}

\DeclareMathOperator{\re}{Re}
\DeclareMathOperator{\im}{Im}
\DeclareMathOperator{\rot}{rot}

\def\revcolor{black}

\def\a{v}
\def\bA{\mathbf{A}}
\def\br{\mathbf{r}}

\def\bk{\mathbf{k}}
\def\bq{\mathbf{q}}
\def\bj{\mathbf{j}}

\newcommand{\corr}[1]{\langle#1\rangle}

\def\jcoeff{\eta}

\newcommand{\aaa}{\tau}

\newcommand{\nb}{\nabla}
\newcommand{\DeltaA}{\Delta_v}
\newcommand{\fzeroq}{\gamma_\text{inh}}

\def\be{\begin{equation}}
\def\ee{\end{equation}}

\begin{document}

\title{Supercurrent flow in inhomogeneous superconductors}

\author{Mikhail A.\ Skvortsov}
\affiliation{L.~D.\ Landau Institute for Theoretical Physics,  Chernogolovka 142432, Russia}
\affiliation{Moscow Institute of Physics and Technology, Dolgoprudny 141701, Russia}

\author{Oleg B.\ Zuev}
\affiliation{L.~D.\ Landau Institute for Theoretical Physics,  Chernogolovka 142432, Russia}
\affiliation{Moscow Institute of Physics and Technology, Dolgoprudny 141701, Russia}

\author{Dina I.\ Fazlizhanova}
\affiliation{HSE University, Moscow 101000, Russia} 

\date{\today}

\begin{abstract}
We study how the supercurrent flow pattern is altered by inhomogeneities in superconducting films. Working in the vicinity of the critical temperature and assuming a model of short-range disorder in the quadratic term of the Ginzburg-Landau functional, we develop a perturbation theory in the inhomogeneity strength. Absorbing the ultraviolet divergences into the renormalization of the critical temperature, we arrive at a well-defined theory governed by large-scale physics.
In the presence of inhomogeneities, the correlation functions of the order parameter and supercurrent exhibit a long-range power-law behavior, which can be attributed to the mixing of the amplitude and phase modes.
The fluctuation magnitude grows with increasing the average current, and the system becomes strongly inhomogeneous near the critical current.
\color{\revcolor}
An inhomogeneous superflow will generate a random magnetic field, whose magnitude in thin NbN films can be comparable to Earth's magnetic field. We discuss the feasibility of detecting it using the SQUID-on-tip technique.
\end{abstract}

\maketitle

\section{Introduction}
\label{sec:one}

The impact of disorder on superconductivity has been the subject of theoretical and experimental research for decades.
According to Anderson's seminal result \cite{Anderson}, the thermodynamic properties of a superconductor in the presence of potential impurity scattering do not depend on the ratio $l/\xi_0$, where $l$ is the mean free path and $\xi_0$ is the clean coherence length.
In particular, the critical temperature, $T_c$, remains the same both in the clean ($l\gg\xi_0$) and dirty ($l\ll\xi_0$) limits, while the coherence length $\xi$ and response functions (superfluid stiffness, optical conductivity \cite{MB,AG59}, etc.) explicitly depend on the degree of disorder.
The insensitivity of $T_c$ to disorder, usually referred to as Anderson's theorem, is valid for good metals with $k_\text{F}l\gg1$, where $k_\text{F}$ is the Fermi momentum. The interplay of disorder and Coulomb interaction weakens the effective attraction in the Cooper channel that results in a suppression of $T_c$ already in the first order in $1/k_\text{F}l$ \cite{Ovchina,Maekawa,Fin87,AntSkv}.
This is the so-called fermionic mechanism of superconductivity suppression \cite{Larkin99}, when the system remains nominally uniform.

Conventional theoretical approaches for describing disordered superconductors, such as the Ginzburg-Landau (GL) expansion \cite{Gor}, Eilenberger \cite{Eilenberger} and Usadel \cite{Usadel} equations, are based (often implicitly) on the conjecture of \emph{self-averaging}. 
The key assumption behind this approximation is that in the presence of homogeneous disorder the superconducting order parameter $\Delta(\br)$ can be replaced by its spatial average. 
Neglecting the emerging inhomogeneity of the order parameter, which is unavoidably introduced by impurities, can be justified for \emph{moderately disordered} superconductors with $k_\text{F}l\gg1$ \cite{SpivakZhou95,FS}. The effect of inhomogeneity becomes dominant for \emph{strongly disordered} superconductors with $k_\text{F}l\sim1$, in the vicinity of the superconductor--insulator transition (SIT) \cite{Fischer,Sacepe}. 
Emerging inhomogeneity of the superconducting state with increasing disorder was demonstrated numerically \cite{Ghosal}. 
\color{\revcolor}
Spontaneous formation of superconducting puddles in a nominally uniform metal opens a way to the bosonic mechanism of superconductivity suppression \cite{Larkin99,Sacepe}. In such a scenario, the macroscopic superconducting state arises due to the interplay of Josephson coupling between the islands and Coulomb repulsion effects \cite{FL98,FSL,PbQSMT}.
\color{black}
Starting with a pioneering experiment \cite{Sacepe-STM}, spontaneous inhomogeneity visualized by the scanning tunneling microscopy/spectroscopy technique has been revealed in a number of disordered superconducting films (TiN, NbN, InO$_x$) \cite{Sacepe-STM2,Chand,Noat,Kamlapure,Cabrilet}, both on the metallic ($k_\text{F}l\gtrsim1$) and insulating ($k_\text{F}l\lesssim1$) sides of the SIT.

\begin{figure*}
\centerline{\includegraphics{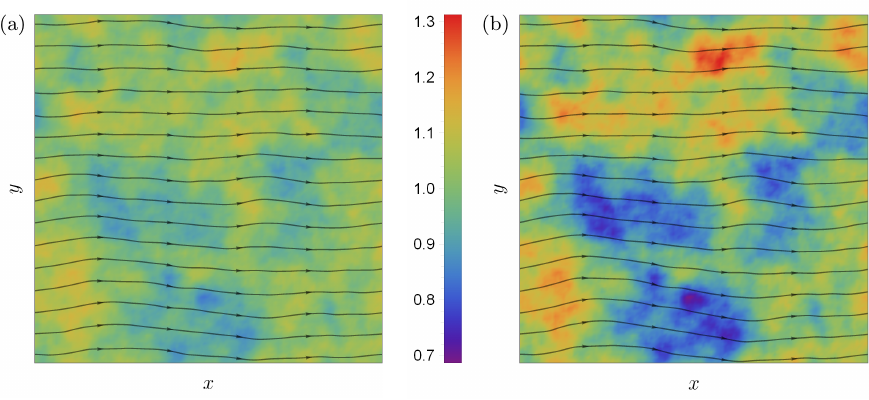}}
\caption{Numerical simulation of the order parameter $|\Delta(\br)|$ (color map; normalized by its value in the homogeneous case) and supercurrent flow $\bj(\br)$ (current flow lines) for an inhomogeneous superconducting film at (a) an infinitesimal current and (b)~close to the critical current, $\a/\a_c = 0.85$. Both panels are obtained for the same short-range disorder with the correlation length $r_c=\xi/10$ in the square domain $12.8\,\xi\times12.8\,\xi$ with periodic boundary conditions. One can clearly see that with increasing the current, the amplitude of $|\Delta(\br)|$ fluctuations grows and their correlation length becomes larger. At the same time, the supercurrent pattern $\bj(\br)$ becomes more inhomogeneous.}
\label{F:modelcurrent}
\end{figure*}

Remarkably, it is this type of disordered films that is now routinely used as the core element of superconducting nanowire single-photon detectors (SNSPD) \cite{Goltsman01,Natarajan12}, as well as in quantum phase-slip devices \cite{Mooij,Astafiev}. 
To make a photon detector out of a superconducting nanowire, it is biased by a finite supercurrent close to the critical current. Upon an arrival of a photon, its energy is released into the electronic system causing local destruction of the supercurrent state followed by a voltage pulse.
Due to an intrinsic inhomogeneity of the superconducting nanowire made of a highly resistive material, the barrier protecting the supercurrent state should exhibit spatial fluctuations. Therefore a photon with a given energy may either trigger a count or not depending on the value of the barrier at the absorption point. This will lead to an effective smearing of the threshold energy of the detector, deteriorating the device performance. Other mechanisms of broadening the sharp spectral cutoff in SNSPDs discussed in literature involve Fano fluctuations (the fraction of energy remaining in the electronic subsystem after the first stage of photon absorption \cite{Fano}) \cite{Kozorezov}, local temperature fluctuations \cite{Semenov}, dependence on the position of the photon absorption point across the strip \cite{Vodolazov}. To the best of our knowledge, the only attempt to describe the effect of spatial inhomogeneity on the operation of SNSPDs was made in Ref.\ \cite{China}, where a simple and rather artificial model of inhomogeneity was used.

The response of a superconductor to an infinitesimally small current is characterized by the superfluid stiffness. Its behavior in superconductors with emerging inhomogeneity was studied numerically in Ref.\ \cite{Seibold12}. Recently, an anomalous power-law temperature dependence of the superfluid stiffness was reported for a strongly disordered amorphous InO$_x$ with localized Cooper pairs ($k_\text{F}l\lesssim1$), which was qualitatively explained by mapping onto a pseudo-spin model on a graph \cite{Khvalyuk24}. 
Despite those attempts, an analytical description of the superfluid stiffness for moderately disordered superconductors on the metallic side of the SIT ($k_\text{F}l\gtrsim1$) is still missing.

In the current publication, we aim at developing a theoretical approach to characterizing the supercurrent state in the presence of inhomogeneity. To simplify the analysis, we restrict ourselves to the vicinity of $T_c$, where the GL expansion in terms of the order parameter field $\Delta(\br)$ can be employed. Since the microscopic origin of inhomogeneity in disordered samples is typically unknown, we will use a phenomenological \emph{random-temperature model}, when the coefficient $\aaa(\br)$ in front of $|\Delta(\br)|^2$ term in the GL functional exhibits short-range spatial fluctuations \cite{IoLar}. This model is believed to be a universal description near $T_c$. It can be derived starting from the random-coupling model \cite{LO-RCC} or from fully microscopic theory of inhomogeneity due to mesoscopic fluctuations \cite{FS}. The emerging inhomogeneity is assumed to be weak, $\mathop{\rm var}|\Delta| \ll \corr{|\Delta|}^2$, and hence can be treated perturbatively starting with a homogeneous supercurrent state. 
At the same time, we neglect thermal fluctuations assuming $(T_c-T)/T_c\gg\text{Gi}$, where $\text{Gi}$ is the Ginzburg~number.

Our main result is the expression for the correlation functions of the order parameter $\Delta(\br)$ and supercurrent $\bj(\br)$ obtained as a function of the \emph{average} dimensionless superfluid velocity $v=\xi\corr{\tilde{\mathbf{v}}_s}$. Figure~\ref{F:modelcurrent} shows the outcome of the numerical simulation for a particular realization of inhomogeneity in the GL coefficient $\aaa(\br)$ at different values of the superfluid velocity $v$. The order parameter modulus shown by color is plotted in units of its value $\DeltaA$ in the homogeneous case given by Eq.~\eqref{Delta0-def}. The left and right panels are obtained for $v\to0$ and $v=0.85\,v_c$, respectively, where $v_c$ is the maximal value of the superfluid velocity in a homogeneous system. Comparison between them clearly demonstrates that the magnitude of spatial fluctuations both of $|\Delta(\br)|$ and $\bj(\br)$ grows with the superfluid velocity. 
These findings are consistent with the exact analytical expressions for the order parameter and supercurrent correlation functions.

The paper is organized as follows. In Sec.\ \ref{sec:two} we introduce the model and discuss the effects of thermal and frozen-in-space fluctuations. Perturbative expansion in the strength of inhomogeneity is developed in Sec.\ \ref{S:pert}, which introduces the basic element of the theory, the fluctuation propagator on top of the uniform supercurrent state. 
The supercurrent correlation function and the correction to its average are calculated in Sec.\ \ref{S:stat}. The results are summarized in Sec.\ \ref{S:concl}. The technical details of calculating the asymptotic behavior of the fluctuation propagator can be found in Appendix \ref{A:L2D}.

\section{Inhomogeneous superconductor near the critical temperature}
\label{sec:two}

\subsection{Ginzburg-Landau functional}

Description of a superconductor near the critical temperature is based on the GL expansion of the free-energy density in powers of the complex order parameter field $\Delta(\br)$. In the presence of inhomogeneity, it can be written in the form
\be
\label{energy}
\frac{F}{\nu}
=
  \xi_0^2 |\nabla\Delta|^2
- [\aaa+\aaa_1(\br)] |\Delta|^2 
+ \frac{\beta}{2} |\Delta|^4 .
\ee
Here $\aaa = (T_{c0}-T)/T_{c0}$, $\beta = 7 \zeta (3)/8\pi^2T_{c0}^2$, and $\xi_0^2=\pi \hbar D/8T_{c0}$ in the dirty-limit considered \cite{Gor} ($\nu$ is the density of states at the Fermi energy and $D$ is the diffusion coefficient).
The dimensionless distance to the transition $\aaa$ is positive in the superconducting phase and vanishes at $T=T_{c0}$, which is the mean-field transition temperature in a homogeneous system.
The constant $\xi_0$ is (up to a numerical factor) the zero-temperature coherence length.
Comparing the quadratic terms in Eq.~\eqref{energy} yields an expression for the temperature-dependent coherence length divergent at the transition:
\be
  \xi(T) = \xi_0/\sqrt{\aaa} = \xi_0\sqrt{T_{c0}/(T_{c0}-T)} .
\ee

Inhomogeneity is naturally introduced as a random variation of the coefficient $\aaa$, while the coefficients $\beta$ and $\xi_0$, which remain finite at the second-order phase transition, are considered constant.
A random field $\aaa_1 (\br)$ is assumed to be Gaussian with the correlation function
\be
\label{f-def}
\corr{\aaa_1(\br) \aaa_1(\br')} = f_\aaa(\br-\br') .
\ee
It is characterized by the strength of local fluctuations, $\corr{\aaa_1^2}=f_\aaa(r=0)$, and the correlation radius $r_c$ determined by the decay of $f_\aaa(r)$.
In what follows we will assume that inhomogeneity in the coefficient $\aaa$ is short range compared to the coherence length:
\be
\label{rc}
  r_c \ll \xi(T) ,
\ee
which is always true near the transition. For example, the mechanism of spontaneous inhomogeneity due to mesoscopic fluctuations leads to $r_c\sim\xi_0$ \cite{FS,FS12}, and the inequality \eqref{rc} holds as long as $T_c-T\ll T_c$ (i.\ e., $\tau\ll1$). For short-range disorder specified by Eq.\ \eqref{rc}, all physical quantities (except for an unobservable shift of $T_c$, see below) depend on the zero Fourier component of the correlation function \eqref{f-def},
\be
\label{f(q=0)}
  \fzeroq 
  =
  f_\tau(q\,{=}\,0)
  =
  \int f_\tau(\br) \, d^dr
  \sim
  \corr{\aaa_1^2} r_c^d .
\ee

The system described by Eq.\ \eqref{energy} is often referred to as the random-temperature model. Note however that for short-correlated disorder in $\tau_1(\br)$ it cannot be interpreted as a random-$T_c$ model. With $r_c\ll\xi(T)$, no regions of well-defined local $T_c(\br)$ exist, and emergent inhomogeneity of $\Delta(\br)$ is a collective effect of many fluctuating domains of size $r_c$.

In the GL free energy density \eqref{energy}, we neglect the contribution of the magnetic field $\textbf{H} = \rot\bA$, both external and induced. Specifically, we assume no external magnetic field and a weak enough magnetic field created by the supercurrent. The latter is true for a thin film (thickness $d_z$) if its width in the direction perpendicular to the current, $w$, is much smaller than the Pearl length $\lambda_\text{Pearl}=\lambda^2/d_z$ \cite{Pearl}, where $\lambda=(\hbar c/e)\sqrt{\beta/32\pi \nu \xi_0^2 \aaa}$ is the bulk penetration depth.
In such a case, screening is ineffective and the current flow (in the absence of inhomogeneity) is uniform across the sample.

In the absence of a magnetic field, one can still introduce a vector potential $\bA$ as a pure gauge, replacing the gradient term in Eq.\ \eqref{energy} by the long derivative:
\be
  |\nabla\Delta|^2
  \to
  \left| \left(\nb + i \bA \right)\Delta \right|^2 .
\ee
The expression for the supercurrent density then reads
\be
\label{current}
\bj = \jcoeff
\left( |\Delta|^2 \bA + \frac{\Delta^* \nb \Delta - \Delta \nb \Delta^*}{2i} \right) ,
\ee
with the coefficient $\jcoeff=4e\nu \xi_0^2/\hbar$. Written in terms of the absolute value and phase of the order parameter, $\Delta = |\Delta|e^{i \phi}$, it takes the form
\be
\label{curr}
\bj = \jcoeff |\Delta|^2 \tilde{\mathbf{v}}_s ,
\ee
where
\be 
  \tilde{\mathbf{v}}_s = \nabla \phi + \bA
\ee
is a gauge-invariant phase gradient proportional to the superfluid velocity. The choice of a particular gauge is a matter of convenience that will be discussed below.

\begin{figure}
\centerline{\includegraphics[width=0.9\columnwidth]{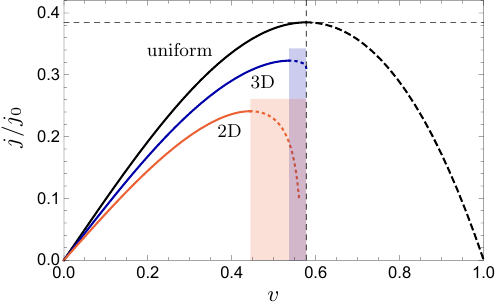}}
\caption{Average supercurrent $\corr{j}/j_0$ vs average dimensionless superfluid velocity. 
Black line: the standard dependence \eqref{j0(a)} in the uniform case. The critical current $j_c/j_0=2/3\sqrt{3}$ is achieved at $v_c=1/\sqrt{3}$, the unstable unphysical branch with $v>v_c$ is shown by the dashed line. 
Blue line: modification in the 3D case given by Eq.\ \eqref{j-av-3D} with $\aaa_*^\text{3D}/\aaa=0.01$.
Red line: modification in the 2D case given by Eq.\ \eqref{j-av-2D} with $\aaa_*^\text{2D}/\aaa=0.05$.
Shaded are fluctuation regions near the bare critical velocity, $v_c-v\lesssim\delta v_d$, with 
$\delta v_3 \sim ( \aaa_*^\text{3D}/\aaa )^{1/3}$ and 
$\delta v_2 \sim ( \aaa_*^\text{2D}/\aaa )^{2/5}$.
The shown first-order perturbative corrections are valid outside of the fluctuation region, at $v_c-v\gtrsim\delta v_d$.
}
\label{finalresults}
\end{figure}

In the mean-field approximation, the order parameter field satisfies the GL equation
\be
\label{GLE}
  -\xi_0^2(\nb+i\bA)^2\Delta
  -[\aaa+\aaa_1(\br)]\Delta
  +\beta|\Delta|^2\Delta=0 .
\ee

\emph{In the absence of inhomogeneity}, a uniform supercurrent suppresses the absolute value of the order parameter $|\Delta|=\DeltaA$ according to 
\be
\label{Delta0-def}
  \DeltaA^2 = \Delta_0^2 (1-v^2) ,
\ee
where $\Delta_0=\sqrt{\aaa/\beta}$ and $v$ is the dimensionless superfluid density,
\be
\label{a-def}
  v= \xi \tilde v_s .
\ee
The current density \eqref{curr} is given by the textbook expression \cite{Tin}
\be
\label{j0(a)}
  j = j_0 (1-v^2) v ,
\ee
where
\be
\label{j0-def}
  j_0 
  = \frac{\aaa \eta}{\beta \xi} 
  = \frac{\aaa^{3/2}\jcoeff}{\beta \xi_0}
  .
\ee
The dependence of the supercurrent on the superfluid velocity given by Eq.\ \eqref{j0(a)} is shown in Fig.\ \ref{finalresults} by the black curve. The critical current is achieved at $v_c=1/\sqrt{3}$. The non-physical branch with $v>v_c$ is marked by the dashed line.

\emph{In the presence of random} $\aaa_1(\br)$, Eq.\ \eqref{GLE} becomes a stochastic nonlinear differential equation. Then the order parameter $\Delta(\br)$ and hence the supercurrent $\bj(\br)$ defined by Eq.\ \eqref{current} also become random quantities. Below we will analyze their statistical properties assuming that spatial fluctuations are weak and can be treated within the lowest order of the perturbation theory. Though the most pertinent application of our theory is related to thin films (two-dimensional case, 2D), we will also discuss the bulk geometry (three-dimensional case, 3D) for pedagogical purposes. The difference between the two limits is mainly in the relation between the film thickness $d_z$ and the coherence length $\xi(T)$.
To be specific, the 2D case refers to $r_c \sim d_z \ll \xi \ll \lambda_\text{Pearl}$, while the 3D case is realized for $r_c \ll \xi \ll d_z \ll \lambda_\text{Pearl}$.

\subsection{Thermal fluctuations vs inhomogeneity}
\label{SS:fluct}

\subsubsection{Thermal fluctuations in a homogeneous medium}
\label{SSS:Thermal}

The GL functional \eqref{energy} with $\aaa_1(\br)=0$ is a paradigmatic model for describing second-order phase transitions in homogeneous systems.
Sufficiently far from the transition the order parameter is well described by the mean-field approximation, while in the vicinity of $T_c$ the increasing role of thermal fluctuations causes a change in the critical behavior \cite{WilsonKogut,Zinn}. The relative width of the temperature region where thermal fluctuations are relevant is characterized by the dimensionless Ginzburg number \cite{Levanyuk,VL-book}
\be
\label{Gi}
  \text{Gi} 
  \sim 
  \left(
    \frac{1}{\nu_d\xi_0^dT_c}
  \right)^{2/(4-d)} ,
\ee
where $d$ is the effective dimensionality of the problem and $\nu_d$ is the corresponding density of states ($\nu_3=\nu$ in 3D, and $\nu_2=\nu d_z$ in 2D, with $d_z$ being the film thickness).

\begin{figure}
    \centering
\includegraphics{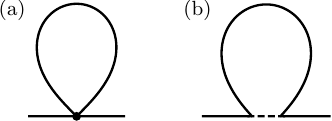}
    \caption{The simplest self-energy diagrams due to (a) thermal fluctuations, $\Sigma^\text{th}$, and (b) inhomogeneity, $\Sigma^\text{dis}$. The dashed line stands for the correlator $\corr{\aaa_1(\br) \aaa_1(\br')}$. For short-range disorder, both diagrams diverge in the ultraviolet in a similar fashion.
}
    \label{F:loop}
\end{figure}

In addition to modification of the critical indices, thermal fluctuations result in a shift of the critical temperature from the bare (and unobservable) value $T_{c0}$ to the actual $T_c$ \cite{VL-book,Cappellaro}. This shift can be obtained from the simple loop diagram for the self-energy shown in Fig.\ \ref{F:loop}(a), which in the normal phase ($\aaa<0$) is given by
\be
\label{Sigma-loop}
  \Sigma^\text{th}
  =
  - 2
  \beta
  \corr{|\Delta(\br)|^2} 
  =
  - \frac{2T}{\nu_d}
  \beta \int \frac{(d\bq)}{|\aaa|+\xi_0^2q^2} ,
\ee
where $(d\bq)\equiv d^dq/(2\pi)^d$.
This momentum integral diverges in the ultraviolet for space dimensions $d\geq 2$. The divergence should be compensated by subtracting a counterterm,
\be
  \Sigma^\text{th}
  \to
  \Sigma^\text{th}
  -
  \Sigma^\text{th}_{\aaa=0} ,
\ee
with a simultaneous renormalization of the coefficient $\aaa$. The shift of the critical temperature is determined by the relation $T_c - T_{c0} = T_{c0}\Sigma^\text{th}(T_c,\bk=0)$. The negative value of the diverging self-energy \eqref{Sigma-loop} indicates that thermal fluctuations suppress $T_c$.

\subsubsection{Phase transition in an inhomogeneous system}
\label{SSS:Inhomo}

The random-temperature model \eqref{energy} with $\aaa_1(\br)\neq 0$ (as well as a somewhat different random-field model) is a
standard playground for studying the effect of quenched disorder on the second-order phase transitions \cite{HarrisLubensky,Khmel,Dotsenko06}.
Here the main focus of theoretical research was understanding the impact of disorder on the critical behavior.
According to the Harris criterion \cite{Harris}, irregularity of the system's parameters is irrelevant if the clean-system 
specific heat index is negative, $\alpha<0$. 
Otherwise, the critical exponents at the phase transition are controlled by disorder modifying the renormalization group flow \cite{HarrisLubensky,Khmel}.

While renormalization group is a sophisticated method of summing perturbative corrections, there exists a nonperturbative effect of disorder related to formation of rare localized islands of the ordered phase above the transition (Griffiths phase). Appearance of such disorder-induced droplets results in a nonanalytic---yet hardly experimentally observable---contribution to the free energy \cite{Griffiths,Dotsenko06}.

Finally, quenched randomness in the GL free energy \eqref{energy} is also responsible for the change of the critical temperature. This shift is determined by the diagram $\Sigma^\text{dis}$ shown in Fig.\ \ref{F:loop}(b), which for a short-range disorder with $r_c\ll\xi$ behaves similar to its analog $\Sigma^\text{th}$ due to thermal fluctuations [see Fig.\ \ref{F:loop}(a)], with the formal replacement $\beta\to-(\nu_d/2T)\fzeroq$, where $\fzeroq$ is the inhomogeneity strength defined in Eq.\ \eqref{f(q=0)}.
Following the reasoning of Sec.\ \ref{SSS:Thermal}, we conclude that the ultraviolet divergency of the self-energy $\Sigma^\text{dis}$ should be absorbed into a redefinition of $T_c$.
Note that while $\Sigma^\text{th}$ is negative, $\Sigma^\text{dis}$ is positive, and therefore the effect of inhomogeneity is to increase the critical temperature.

The relative strength of quenched disorder is characterized by the dimensionless number [for the exact definitions, see Eqs.\ \eqref{alpha*3D} and \eqref{alpha*2D}]
\be
\label{alpha*d}
  \aaa_*
  \sim
  \left(
  \frac{\fzeroq}{\xi_0^d}
  \right)^{2/(4-d)} ,
\ee
with $\fzeroq$ given by Eq.\ \eqref{f(q=0)}. For inhomogeneous systems, the parameter $\aaa_*$ plays the role of the Ginzburg number \eqref{Gi} in the theory of thermal fluctuations. Both determine the width of the temperature regions near $T_c$, where frozen-in-space/thermal fluctuations become strong. The mean-field description is valid as long as 
\be
  \tau\gg\max(\text{Gi},\tau_*).
\ee
In what follows we assume that this inequality holds and consider the effect of a weak inhomogeneity in the leading order in $\tau_*/\tau$. Since $\tau\gg\text{Gi}$, thermal fluctuations will not wash out the effect of inhomogeneity.

\section{Perturbation theory}
\label{S:pert}

\subsection{Recursive determination of $\Delta$}

The nonlinear stochastic GL equation \eqref{GLE} can be formally solved by perturbative series expansion in $\aaa_1(\br)$ on top of the uniform supercurrent state. To describe the latter, we choose the real gauge with $\phi=0$, $\bA=\tilde{\mathbf{v}}_s$ and $\Delta=\DeltaA$, as given by Eq.\ \eqref{Delta0-def}. For a finite $\tau_1(\br)$, we write the order parameter as
\be
\label{Delta-exp}
  \Delta(\br) = \DeltaA + \Delta_1(\br) + \Delta_2(\br) + \dots , 
\ee
where $\Delta_n$ scales as the $n$th power of $\aaa_1$. In the presence of a supercurrent, inhomogeneities affect both the absolute value and the phase of the order parameter, so $\Delta_n(\br)$ are complex numbers. It is convenient to organize $\Delta_n$ and $\Delta_n^*$ into a 2-vector,  which will enable us to formulate a theory without the use of the complex conjugation operation.

Substituting the expansion \eqref{Delta-exp} into the GL equation \eqref{GLE} and its conjugate, and balancing equal powers of $\aaa_1$, we arrive at a set of recursive relations:
\be
\label{recursive}
  L^{-1}
  \begin{pmatrix}
    \Delta_n(\br) \\ \Delta_n^*(\br)
  \end{pmatrix}
  =
  \Pi_n(\br) ,
\ee
where $L^{-1}$ is a differential matrix operator
\be
\label{L^-1-def}
  L^{-1}
  =
  - \xi_0^2(\sigma_0\nb^2+2i\bA\sigma_3\nb) + \beta\DeltaA^2 (\sigma_0+\sigma_1)
  ,
\ee
$\sigma_i$ being the Pauli matrices.
The vectors $\Pi_n(\br)$ contain only $\aaa_1(\br)$ and $\Delta_k(\br)$ with $k<n$.
The first two vectors are given by
\begin{subequations}
\begin{gather}
\label{Pi1}
  \Pi_1
  =
  \DeltaA 
  \begin{pmatrix}
    \aaa_1 \\ \aaa_1
  \end{pmatrix} ,
\\
\label{Pi2}
  \Pi_2
  =
  \begin{pmatrix}
    \aaa_1 \Delta_1 
    - \beta \DeltaA (\Delta_1^2+2\Delta_1\Delta_1^*)
  \\
    \aaa_1 \Delta_1^* 
    - \beta \DeltaA (\Delta_1^{*2}+2\Delta_1\Delta_1^*)
  \end{pmatrix} .
\end{gather}
\end{subequations}

Solving equations \eqref{recursive} step by step, one can determine all terms in the perturbative expansion \eqref{Delta-exp} in powers of $\aaa_1$ for a given inhomogeneity realization. 
Graphically, this procedure corresponds to the summation of tree-like diagrams. The absence of closed loops [as in Fig.\ \ref{F:loop}(a)] corresponds to the neglect of thermal fluctuations.

\subsection{Fluctuation propagator}
\label{SS:FP}

The key ingredient of the theory is the fluctuation propagator $L$, the resolvent of the differential operator in Eq.\ \eqref{L^-1-def}. (A conventional definition of the fluctuation propagator contains the factor of $T$ \cite{VL-book}. On a tree level considered, it is completely compensated by the factors $1/T$ from the vertices in the Gibbs weight.)
In real space, $L(\br,\br')$ is an integral kernel depending on the coordinate difference. In the momentum space, it takes the form
\be
\label{L(q)}
  L(\bq)
  =\frac{
  \xi_0^2(\bq^2\sigma_0-2\bA\bq\sigma_3) 
  + \beta \DeltaA^2 (\sigma_0-\sigma_1)
}{D(\bq)} ,
\ee
with $D(\bq) = \det L^{-1}(\bq)$ given by
\be
  D(\bq)
  = 
  \xi_0^4[q^4-4(\bA\bq)^2] + 2 \beta \DeltaA^2 \xi_0^2 q^2 .
\ee
It is convenient to introduce the dimensionless momentum normalized by the temperature-dependent coherence length:
\be
\label{q-dimless}
\bq' = \xi \bq .
\ee
Then the function $D(\bq)$ acquires the form:
\be
  D(\bq)
  = 
  \aaa^2[q'^4 + 2 (1-v^2) q'^2 - 4v^2q_x'^2] ,
\ee
where $v=\xi A$ is the dimensionless superfluid velocity introduced in Eq.\ \eqref{a-def}, and the current $\bj\parallel\bA$ is assumed to flow in the $x$ direction. 

In the absence of a supercurrent ($\bA=0$), the determinant factorizes: $D(\bq) = \xi_0^2 q^2(\xi_0^2 q^2 + 2\aaa)$. That corresponds to decoupling (at the Gaussian level) of massive fluctuations of the order parameter modulus 
and massless fluctuations of the order parameter phase 
(Goldstone mode), with the isotropic propagators given by, respectively, 
\be
\label{modes}
  L_\parallel^{(0)}(\bq) = \frac{1}{\xi_0^2 q^2 + 2\aaa} ,
  \qquad
  L_\perp^{(0)}(\bq) = \frac{1}{\xi_0^2 q^2} .
\ee
In the coordinate representation, $L_\parallel^{(0)}(\br)$ is the Yukawa potential:
\be
\label{L-par-A0}
  L_\parallel^{(0)}(\br)
  =
  \begin{cases}
    K_0(\sqrt{2}r/\xi)/2\pi \xi_0^2 , & \text{2D} ,
    \\[3pt]
    \exp(-\sqrt{2}r/\xi)/4\pi \xi_0^2 r , & \text{3D} , 
  \end{cases}
\ee
where $K_0$ is the modified Bessel function.

The fluctuation propagator on top of a uniform supercurrent state with $\bA\neq0$ describes a nontrivial mixing of the amplitude and phase modes. Physically, this mixing is a consequence of the current conservation respected by the GL equation. Therefore, in a supercurrent state any change of the order parameter modulus should be compensated by an appropriate  change of its phase to ensure a divergence-free current \cite{Seibold12}. The fluctuation propagator in the one-dimensional geometry (current-carrying superconducting wire) were studied in Ref.\ \cite{McCumber} in order to describe fluctuations around the Langer-Ambegaokar instanton \cite{LA}.

The mixing of the amplitude and phase modes can be seen in the eigenvalues of the matrix $L^{-1}(\bq)$, which are given by 
\be
\label{Lambda}
  \Lambda_\pm(\bq)
  =
  \aaa
  \left[
  1-\a^2 + q'^2
  \pm
  \sqrt{(1-\a^2)^2 + 4\a^2q_x'^2}
  \right] .
\ee
In the absence of a supercurrent, $\Lambda_+(\bq)$ and $\Lambda_-(\bq)$ correspond to the amplitude and phase modes, respectively.
For a finite $\bA$, $\Lambda_\pm(\bq)$ can no longer be interpreted as energies of appropriately gapped parabolic bands, in contrast to Eq.~\eqref{modes}. In addition, the $\bq$ dependence of the eigenvalues $\Lambda_\pm(\bq)$ becomes essentially anisotropic.

The lowest eigenvalue, $\Lambda_-(\bq)$, describes a Goldstone mode with $\Lambda_-(0)=0$. Its behavior is fundamentally different for $\a<\a_c$ and $\a>\a_c$. At the physical branch ($\a<\a_c$), $\Lambda_-(\bq)>0$ for all finite momenta. Whereas at the unphysical branch  ($\a>\a_c$), the function $\Lambda_-(\bq)$ becomes negative in some region of the momentum space in the vicinity of $\bq=0$ that signals instability of the unphysical branch.

\subsection{Modulus response}
\label{SS:mod-resp}

To visualize the spatial structure of the fluctuation propagator, it is instructive to look at the linear combination of its four elements,
\be
\label{LLLL}
  {\cal L} = (L_{11}+L_{12}+L_{21}+L_{22})/2 ,
\ee
which determines the response of the order parameter \emph{modulus} to spatial variations of the GL coefficient $\aaa$ [see Eqs.\ \eqref{Delta1} and \eqref{M-def} below]. In the momentum representation, it reads 
\be
\label{cal-L(q)}
  {\cal L}(\bq)
  =
  \frac{\xi_0^2q^2}{D(\bq)}
  =
  \frac{q'^2}{\aaa[q'^4 + 2 (1-\a^2) q'^2 - 4\a^2q_x'^2]}
  ,
\ee
with the dimensionless momentum $q'$ defined in Eq.\ \eqref{q-dimless}.

\begin{figure}
\centerline{\includegraphics[width=1\columnwidth]{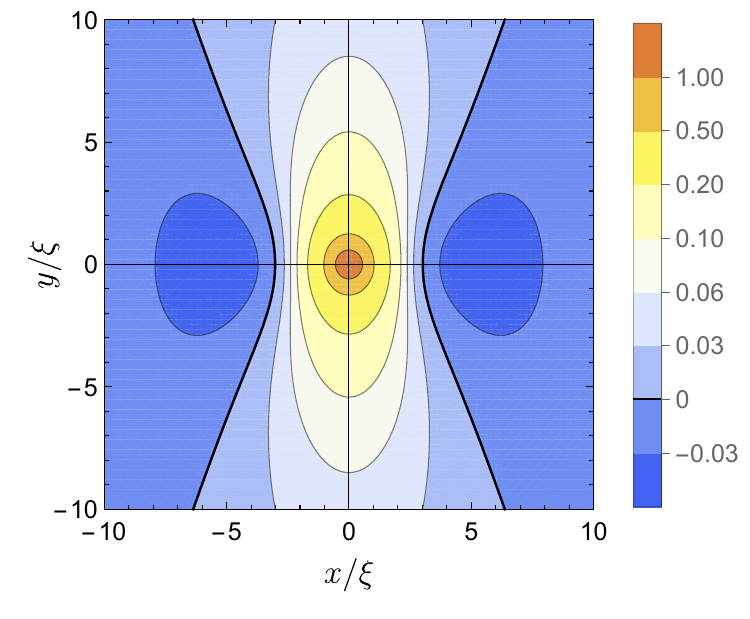}}
\caption{Spatial dependence of the 2D fluctuation propagator ${\cal{L}}(\br)$ (in units of $1/2\pi\aaa \xi^2$) for $\a/\a_c=0.95$. The isotropic peak at $r \ll \xi$, ${\cal{L}}(\br) \approx \ln(1/r)/2\pi$, crosses over to an anisotropic power-law decay \eqref{L-far} at $r\gg\xi$.}
\label{F:LLLL}
\end{figure}

Figure \ref{F:LLLL} shows ${\cal L}(\br)$ in the 2D geometry for the superfluid velocity close to the critical one, $\a/\a_c=0.95$.
At the smallest scales, $r\ll\xi$, the fluctuation propagator $\cal L(\br)$ follows the zero-current limit of the amplitude mode $L_\parallel^{(0)}(\br)$ given by Eq.\ \eqref{L-par-A0}. 

A new feature arising in the presence of a supercurrent is that in the limit of large $\br$ the fluctuation propagator does not decay exponentially, but exhibits a much slower power-law dependence with the exponent equal to the space dimensionality $d$:
\be
\label{L-far}
  {\cal L}^\text{far}(\br) 
  \approx 
  \frac{\a^2}{2\pi\aaa} \frac{\lambda_d(\a, \theta)}{r^d} ,
\ee
where $\theta$ is the angle the vector $\br$ makes with the $x$-axes (direction of $\bA$). The power-law behavior of the modulus response $\cal L(\br)$ is a consequence of mixing between the massive and massless modes of the zero-current fluctuation propagators [Eq.\ \eqref{modes}] at a finite~$\bA$.

The asymptotic behavior \eqref{L-far} is obtained by neglecting the quartic term in the denominator of Eq.\ \eqref{cal-L(q)}, writing it through the Laplace transform, taking the Gaussian integral over $\bq$, and performing the final elementary integration. This way we get in the 2D and 3D cases:
\begin{gather}
\label{lambda2}
  \lambda_2 
  = 
  \frac{1-3\a^2-2(1-2\a^2)\cos^2\theta}
  {\sqrt{(1-\a^2)(1-3\a^2)}(1-3\a^2+2\a^2\cos^2\theta)^2} ,
\\
\lambda_3
  = 
  \frac{1-3\a^2-(3-5\a^2)\cos^2\theta}
  {2\sqrt{1-\a^2}(1-3\a^2+2\a^2\cos^2\theta)^{5/2}} .
\end{gather}
Both functions are negative for $0\leq\theta<\theta_d(\a)$ and positive for $\theta_d(\a)<\theta\leq\pi/2$ [see Fig.\ \ref{F:LLLL}]. The angle $\theta_d(\a)$ separating the two regimes is given by ($d=2$, 3)
\be
  \cos^2\theta_d(\a) = (1-3\a^2) / [d-(d+2)\a^2] .
\ee
It approaches $\pi/2$ at criticality, $\a\to \a_c$, indicating that in the far asymptotics \eqref{L-far} the propagator is negative everywhere except for a small sector near the $y$ axes [for the parameters of Fig.\ \ref{F:LLLL}, $\theta_2(a)=0.77\,\pi/2$]. At the same time, the narrow peak of ${\cal L}^\text{far}(\br)$ in the direction perpendicular to the current is much higher than its behavior along the current, $\lambda_d(\pi/2)/|\lambda_d(0)| \sim (\a_c-\a)^{-d/2}$.

\emph{Right at the criticality} ($\a=\a_c$), the fluctuation propagator ${\cal L}(\br)$ in the 2D geometry can be evaluated in a closed form. To this end we note that its Fourier transform may be written as ${\cal L}(\bq) = \re\aaa^{-1}/(q'^2+2iq'_y/\sqrt{3})$. Writing it through the Laplace transform renders $\bq$ integration Gaussian, and after some algebra we arrive at
\be
\label{exact2D}
  {\cal{L}}_\text{2D}^\text{crit}(\br) 
  = 
  \frac{1}{2\pi\aaa\xi^2} K_0\left(\sqrt{\frac{x^2+y^2}{3\xi^2}}\right) \cosh\frac{y}{\sqrt{3}\xi} .
\ee
Remarkably, its dependence \emph{along} the current ($x$ axis) coincides---up to coordinate rescaling by the factor of $\sqrt6$---with that at $\bA=0$ [Eq.\ \eqref{L-par-A0}], showing an exponential decay at $x\gg\xi$. However, the decay of correlations \emph{perpendicular} to the current ($y$ axis) is only power-law: ${\cal{L}}_\text{2D}^\text{crit}(0,y)\propto y^{-1/2}$.

In the 3D case the propagator at the criticality cannot be obtained in a closed form. Its log-distance tail contains a messy combination of hypergeometric functions, so we present only its asymptotic behavior along the main axes:
\be
  {\cal{L}}_\text{3D}^\text{crit}(\br) 
  \approx
  \begin{cases}
    \displaystyle
    - \frac{3}{4\pi\aaa}\frac{1}{x^3} , & x\to\infty , 
  \\[12pt]
    \displaystyle
    \frac{0.025}{\aaa\xi^{3/2}} \frac{1}{y^{3/2}} , & y\to\infty .
  \end{cases}
\label{critical3D}
\ee
[The number in the last line stands for the exact expression $3^{1/4}\Gamma(3/4)/2^{5/2}\pi\Gamma(1/4)$.]

To conclude this Section let us discuss how and when various asymptotic regimes mentioned above transform into each other.

For \emph{small currents} ($\a\ll \a_c$), the propagator ${\cal L}(\br)$ follows the Yukawa potential \eqref{L-par-A0} up to $r/\xi\sim\ln(1/\a)\gg1$, where its exponential decay crosses over to the power-law behavior ${\cal L}^\text{far}(\br)$ given by Eq.\ \eqref{L-far}.

For \emph{nearly critical currents} ($\a_c-\a\ll \a_c$), the propagator follows it critical asymptotics ${\cal L}^\text{crit}(\br)$ at small $r$ and far asymptotics ${\cal L}^\text{far}(\br)$ at large $r$.
In the 3D geometry, there is a direct crossover between these regimes taking place at
\be
\label{condition}
  r_c(\theta) 
  \sim 
  \xi
  \left[ 
    (\a_c-\a)\cos^2\theta + (\a_c-\a)^2
  \right]^{-1/2} 
  \gg \xi
\ee
[note that different scaling of $r_c(\theta)$ with $\a_c-\a$ along the $x$ and $y$ axes is consistent with different power-law asymptotics~\eqref{critical3D}].
On the contrary, in the 2D case, a new intermediate asymptotic region appears between the small- and large-$r$ types of behavior, where the fluctuation propagator is nearly constant:
\be
\label{L2Dint}
  {\cal L}_\text{2D}^\text{int}(\br) 
  \approx
  - \frac{3^{3/4}\sqrt{\a_c-\a}}{2\pi\aaa\xi^2}.
\ee
This new regime is realized at $r_2(\theta)\lesssim r\ll r_c(\theta)$, where
\be
\label{condition2}
  r_2(\theta) \sim \frac{\sqrt{3}\,\xi}{2(1-|\sin\theta|)} \ln\frac1{\a_c-\a}
  \gg \xi .
\ee
In narrow sectors around $\theta=\pm\pi/2$, where $r_2(\theta)>r_c(\theta)$, there is a direct crossover between ${\cal L}^\text{crit}(\br)$ and ${\cal L}^\text{far}(\br)$ taking place at $r_c(\theta)$. Derivation of Eq.\ \eqref{L2Dint} is presented in Appendix \ref{A:L2D}.

\subsection{Lowest-order corrections $\Delta_1$ and $\Delta_2$}

Now we proceed to the step-by-step determination of the terms $\Delta_n$ in the perturbative expansion \eqref{Delta-exp} in powers of $\aaa_1$.
Solving Eq.\ \eqref{recursive} for $n=1$ with $\Pi_1$ given by Eq.\ \eqref{Pi1}, we obtain for the first correction:
\be
\label{Delta1}
  \begin{pmatrix}
    \Delta_1(\bq) \\ \Delta_1^*(\bq)
  \end{pmatrix}
  =
  \DeltaA
  \begin{pmatrix}
    M(\bq) \\
    M(-\bq)
  \end{pmatrix}
  \aaa_1(\bq) ,
\ee
where $M(\bq)=L_{11}(\bq)+L_{12}(\bq)=L_{21}(-\bq)+L_{22}(-\bq)$ is given by
\be
\label{M-def}
  M(\bq) = \frac{\xi_0^2(q^2-2\bA\bq)}{D(\bq)} .
\ee
Since $M(\bq)\propto 1/q^2$ for large $q$, all correlation functions of $\Delta_1$ and $\Delta_1^*$ converge in the ultraviolet. The response of the order parameter modulus discussed in Sec.\ \ref{SS:mod-resp} is governed by ${\cal L}(\bq)=[M(\bq)+M(-\bq)]/2$, see Eqs.\ \eqref{LLLL} and \eqref{cal-L(q)}.

In a similar way, the quadratic-in-$\aaa_1$ correction, $\Delta_2(\br)$, can be obtained from the recursive equation \eqref{recursive}, with $\Pi_2$ given by Eq.\ \eqref{Pi2} and $\Delta_1$ taken from Eq.\ \eqref{Delta1}.
Since we are going to study statistics of the order parameter and supercurrent in the lowest order in inhomogeneity, we will need only the average value of the second correction, $\corr{\Delta_2(\br)}$. 
The latter is determined by Eq.\ \eqref{recursive} with the space-independent $\corr{\Pi_2(\br)}$ in the right-hand side. The resulting equation for $\corr{\Delta_2}$ should be solved with care. An attempt to do that in the momentum space by inverting $L^{-1}$ fails since $L(\bq)$ is ill-defined at $\bq=0$.
Instead, we will look at Eq.\ \eqref{recursive} for $\corr{\Delta_2(\br)}$ in real space, that allows us to clarify the origin of the above ambiguity.

First, we note that the two components of the vector $\corr{\Pi_2}$ are equal to each other. Second, we restrict $\corr{\Delta_2(\br)}$ and $\corr{\Delta_2^*(\br)}$ to the class of functions that are constant or linear in $\br$, since other choices contradict space uniformity. Neglecting then the second derivative, we present Eq.\ \eqref{recursive} for $\corr{\Delta_2(\br)}$ in the form
\begin{subequations}
\begin{gather}
  \bA \nb \corr{\re \Delta_2(\br)} = 0 ,
\\
  \xi_0^2 \bA \nb \corr{\im \Delta_2(\br)}
  + \beta\DeltaA^2 \corr{\re \Delta_2(\br)}
  =
  \corr{\Pi_2}/2 .
\end{gather}
\end{subequations}
This system of equations tells us that $\corr{\re\Delta_2}$ is constant in space, but does not fix its value that depends on $\nb\corr{\im \Delta_2(\br)}$, which has the meaning of an additional phase gradient of the order parameter field.

Hence the ambiguity in the determination of $\corr{\Delta_2(\br)}$ is related to the fact that a brute-force perturbation theory in $\aaa_1$ leaves the average value of the superfluid velocity unspecified, as long as no boundary condition on the phase is imposed. The boundary condition is our freedom, and we choose it by requiring that the average superfluid velocity $\corr{\tilde{\mathbf{v}}_s}$ is not renormalized in the presence of inhomogeneity and is still given by $\bA$. In terms of the average second-order correction, this means
\be
\label{d2=d2=}
  \corr{\Delta_2}
  =
  \corr{\Delta_2^*}
  =
  \corr{\Pi_2}/2\beta\DeltaA^2 .
\ee
The same reasoning can be applied to higher-order terms $\Delta_{2n}$. As a result, we arrive at a theory, where the supercurrent is obtained perturbatively for a given value of the average superfluid velocity $\corr{\tilde{\mathbf{v}}_s}=\bA$.

\subsection{Average $|\Delta|^2$}
\label{av-Delta2}

The average value of $|\Delta(\br)|^2$ is given by 
\be
  \corr{|\Delta|^2}
  =
  \DeltaA^2
  +\DeltaA \corr{\Delta_2+\Delta_2^*}
  + \corr{\Delta_1^*\Delta_1}
  + \dots ,
\ee
where $\DeltaA$ is the order parameter in the presence of the supercurrent, see Eq.\ \eqref{Delta0-def}.
With the help of Eqs.\ \eqref{Pi2} and \eqref{d2=d2=} we obtain in the lowest order in $f_\aaa$:
\be
\label{d2c}
  \corr{|\Delta|^2}
  =
  \DeltaA^2
  + \corr{\aaa_1 \Delta_1}/\beta\DeltaA 
    - \corr{\Delta_1^2}
  - \corr{\Delta_1^*\Delta_1}
  + \dots
\ee

While the last two terms in Eq.\ \eqref{d2c} are determined by large scales [see discussion below Eq.\ \eqref{M-def}], the term $\corr{\aaa_1 \Delta_1}$ diverges in the ultraviolet for short-range inhomogeneities with $r_c\ll\xi$. 
It is this term, which is responsible for the shift of $T_c$ by quenched disorder in the GL coefficient $\aaa$. As explained in Sec.\ \ref{SSS:Inhomo}, the effect of fluctuations on the critical temperature is taken into account by simultaneously (i) replacing the bare and unobservable value of $T_{c0}$ with the actual $T_c$ and (ii) regularizing all ultraviolet correlation functions according to
\be
\label{ad-reg}
  \corr{\aaa_1 \Delta_1}
  \to
  \corr{\aaa_1 \Delta_1}^\text{reg}
  \equiv
  \corr{\aaa_1 \Delta_1}
  -
  \corr{\aaa_1 \Delta_1}_{\bA=0,\aaa=0} .
\ee
Using Eq.\ \eqref{Delta1} and averaging with the help of Eq.\ \eqref{f-def}, we obtain
\begin{align}
\nonumber
\!  \corr{\aaa_1 \Delta_1}^\text{reg}
  & = 
  \DeltaA 
  \int (d\bq) f_\aaa(\bq) 
  \left[
    \frac{\xi_0^2q^2}{D(\bq)}
    -
    \frac{1}{\xi_0^2q^2}
  \right] 
\\{}
\label{reg}
  & = 
  \fzeroq \DeltaA 
  \int (d\bq) 
    \frac{4\xi_0^2(\bA\bq)^2 - 2 \beta \DeltaA^2 q^2
}{q^2D(\bq)} .
\end{align}

In what follows it will be assumed that the regularization \eqref{reg} has been performed. After subtracting the counterterm, the momentum integration converges in the ultraviolet, and therefore we have replaced $f_\aaa(\bq)$ by $\fzeroq=f_\aaa(q\,{=}\,0)$ and taken it out of the integral sign. As there are no other ultraviolet divergences in the theory, $\fzeroq$ is the only attribute of disorder that will appear hereafter.

As a result, we obtain for the average square of the modulus of the order parameter:
\be
\label{Delta-sq-av}
  \corr{|\Delta|^2}
  =
  \DeltaA^2
  + 
  \frac{\fzeroq}{\beta}
  \int (d\bq)
  \biggl[
    \frac{\xi_0^2q^2}{D(\bq)}
    -
    \frac{1}{\xi_0^2q^2}
    - 
    \frac{2\beta \DeltaA^2 \xi_0^4q^4}{D^2(\bq)}
  \biggr] .
\ee

In order to illustrate the effect of inhomogeneity it is instructive to study $\corr{|\Delta(\br)|^2}$ in the absence of a supercurrent ($\bA=0$):
\be
\label{<DD>-A0}
  \corr{|\Delta|^2}_{\bA=0}
  =
  \frac{\aaa}{\beta}
  \left[
  1
  - 
  4 \fzeroq
  \int
    \frac{(d\bq)(\xi_0^2 q^2 + \aaa)}{\xi_0^2 q^2(\xi_0^2 q^2 + 2\aaa)^2}
  \right] .
\ee

\emph{In the 3D case}, the integral is determined by $q\sim1/\xi$ (the relevant spatial scale is $\xi$), and we obtain
\be
\label{<DD>-A0-3D}
  \frac{\corr{|\Delta|^2}_{\bA=0}^\text{3D}}{\aaa/\beta}
  =
  1
  - 
  \sqrt{\frac{\aaa_*^\text{3D}}{\aaa}}
  ,
\ee
where we have defined
\be
\label{alpha*3D}
  \aaa_*^\text{3D}
  =
  \left(
  \frac{3\fzeroq}{2^{5/2}\pi\xi_0^3}
  \right)^2 ,
\ee
in accordance with Eq.\ \eqref{alpha*d}.

\emph{The 2D case} is marginal. Here the momentum integral in Eq.\ \eqref{<DD>-A0} logarithmically diverges in the infrared [which is an artifact of the ultraviolet regularization \eqref{ad-reg}] and should be cut off at some large scale $L_*$. The proper choice of $L_*$ has been discussed both in the context of thermal fluctuations in homogeneous systems \cite{VL-book,Cappellaro} and frozen-in-space fluctuations in inhomogeneous systems \cite{FS12}. In both cases, the cutoff scale $L_*$ is identified with the coherence length at the border of the fluctuation region, at $\delta T_c/T_c$ equal to either Gi or $\aaa_*$, respectively. Thereby we obtain 
\be
\label{<DD>-A0-2D}
  \frac{\corr{|\Delta|^2}_{\bA=0}^\text{2D}}{\aaa/\beta}
  =
  1
  - 
  \frac{\aaa_*^\text{2D}}{\aaa}
  \ln\frac{\aaa}{\aaa_*^\text{2D}} ,
\ee
where the transition smearing due to inhomogeneity is
\be
\label{alpha*2D}
  \aaa_*^\text{2D}
  =
  \frac{\fzeroq}{4 \pi \xi_0^2} ,
\ee
providing a numerical coefficient to Eq.\ \eqref{alpha*d}.
Equation \eqref{<DD>-A0-2D} written with logarithmic accuracy is valid outside the fluctuation region, at $\aaa\gg\aaa_*^\text{2D}$.
 
Equations \eqref{<DD>-A0-3D} and \eqref{<DD>-A0-2D} demonstrate that frozen-in-space fluctuations of $\Delta(\br)$ induced by quenched inhomogeneity in the GL coefficient $\aaa$ lead to the suppression of the average $\corr{|\Delta(\br)|^2}$ compared to its mean-field value $\aaa/\beta$ (after renormalizing $T_c$). These fluctuations become strong at $\aaa\lesssim\aaa_*$ that changes the nature of the transition and requires a nonperturbative treatment. Only significantly below the (renormalized) $T_c$, at $\aaa\gg\aaa_*$, spatial fluctuations are weak, and their perturbative effect on the average $\corr{|\Delta(\br)|^2}$ is given by Eq.~\eqref{<DD>-A0-3D} and \eqref{<DD>-A0-2D}.

\section{Supercurrent statistics}
\label{S:stat}

\subsection{Average supercurrent}

Now we proceed to calculating the average supercurrent induced by the average superfluid velocity $\corr{\tilde{\mathbf{v}}_s}=\bA$. According to Eq.\ \eqref{current}, the current is a sum of two contributions, $\bj=\bj_1+\bj_2$. Averaging the first term produces $\corr{\bj_1}=\jcoeff\corr{|\Delta|^2} \bA$, where $\corr{|\Delta|^2}$ is given by Eq.\ \eqref{Delta-sq-av}. For the second term we obtain
\be
  \corr{\bj_2}
  =
  \jcoeff
  \DeltaA^2 \fzeroq \int (d\bq)
  \bq M^2(\bq) .
\ee
Substituting $M(\bq)$ from Eq.\ \eqref{M-def}, we see that the resulting vector is aligned along $\bA$ and arrive at
\be
\label{j2-fin}
  \corr{\bj_2}
  =
  - 4
  \jcoeff
  \DeltaA^2 \fzeroq 
  \bA \int (d\bq)
  \frac{\xi_0^4 q^2 q_x^2}{D^2(\bq)}
  .
\ee
Adding the two contributions, we come to the following expression for the average supercurrent:
\be
\label{j-av-1}
  \corr{\bj}
  =
  \jcoeff 
  \left[
  \DeltaA^2
  + 
  \frac{\fzeroq}{\beta}
  \int (d\bq) K(\bq)
  \right] 
  \bA ,
\ee
where 
\be
  K(\bq)
  = 
  \frac{\xi_0^2q^2}{D(\bq)}
  -
  \frac{1}{\xi_0^2q^2}
  - 
  \frac{2\beta \DeltaA^2\xi_0^4q^2(q^2+2q_x^2)}{D^2(\bq)}
  .
\ee

Equation \eqref{j-av-1} describes the lowest-order perturbative correction to the average current, which takes into account both modification of the order-parameter modulus and redistribution of its phase.

\subsubsection{Average current in 3D}

In the 3D geometry, the momentum integral \eqref{j-av-1} is easily evaluated in spherical coordinates and we obtain the following correction to the standard homogeneous relation \eqref{j0(a)}:
\be
\label{j-av-3D}
  \corr{j}
  =
  j_0
  \Biggl[
  1-\a^2 
  + \sqrt{\frac{\aaa_*^\text{3D}}{\aaa}} R_3(\a)
  \Biggr] \a ,
\ee
where the parameter $\aaa_*^\text{3D}$ quantifying the inhomogeneity strength is introduced in Eq.\ \eqref{alpha*3D}, and the dimensionless function $R_3(\a)$ is given by
\be
\label{R3}
  R_3
  =
  \frac{(1-3\a^2)^{3/2} }{6 \a^2}
  -
  \frac{(1-\a^2)(1+3\a^2)
     \arcsin{\sqrt{\frac{2\a^2}{1-\a^2}}}
  }{6\sqrt{2} \a^3} .
\ee

The function $R_3(a)$ is negative for all physically available phase gradients, $0<\a<\a_c$, indicating the suppression of the average current by quenched inhomogeneities. In the limit of small $\a$, one finds $R_3(\a)\approx-11/9$, which translates into the suppression of the superfluid density:
\be
  \frac{\delta n_s}{n_s} 
  = 
  - \frac{11}{9} \sqrt{\frac{\aaa_*^\text{3D}}{\aaa}} .
\ee
At $\a\to\a_c$, the function $R_3(\a)$ remains finite but exhibits a non-analytic square-root behavior:
\be
\label{R3sqrt}
  R_3(\a)\approx -\pi/\sqrt6 + 2^{1/2}3^{1/4}\sqrt{\a_c-\a} .
\ee

The dependence of the average current as a function of the average superfluid velocity in the 3D case obtained from Eqs.\ \eqref{j-av-3D} and \eqref{R3} for $\aaa_*^\text{3D}/\aaa = 0.01$ is shown in Fig.~\ref{finalresults} by the blue line.

\subsubsection{Average current in 2D}

In the 2D geometry, the momentum integral \eqref{j-av-1} is easily evaluated in polar coordinates. Cutting the integral logarithmically divergent in the infrared at $L_*$ as discussed in Sec.\ \ref{av-Delta2}, we obtain
\be
\label{j-av-2D}
  \corr{j}
  =
  j_0
  \left[ 1-\a^2
  + \frac{\aaa_*^\text{2D}}{\aaa} R_2(\a) \right] \a ,
\ee
where the parameter $\aaa_*^\text{2D}$ is defined in Eq.\ \eqref{alpha*2D} and the dimensionless function $R_2(\a)$ is given by
\begin{multline}
\label{R2}
  R_2(\a) 
  = 
  - \ln\frac{\aaa}{\aaa_*^\text{2D}}
  - 2\ln\frac{\sqrt{1-\a^2}+\sqrt{1-3\a^2}}{2} 
\\{}
  - \frac{1}{\a^2}\left(\frac{\sqrt{1-\a^2}}{\sqrt{1-3\a^2}}-1\right)
  .
\end{multline}
In this formula, the contribution of $\corr{j_2}$ [see Eq.\ \eqref{j2-fin}] is $-\sqrt{1-\a^2}/\sqrt{1-3\a^2}$, while the rest is the contribution of $\corr{j_1}$ proportional to $\corr{|\Delta|^2}$. The latter reduces to \eqref{<DD>-A0-2D} at vanishing current.

Like in the 3D case, the function $R_2(\a)$ is negative for $0<\a<\a_c$. In the limit of small $\a$, with logarithmic accuracy $R_2(\a)$ is determined by the first term, and for the suppression of the superfluid density we obtain
\be
\label{dns2D}
  \frac{\delta n_s}{n_s} 
  \approx 
  - \frac{\aaa_*^\text{2D}}{\aaa} \ln\frac{\aaa}{\aaa_*^\text{2D}} .
\ee
In the vicinity of the bare critical current, at $\a\to \a_c$, the function $R_2(\a)$ diverges in a square-root manner:
\be
\label{R2sqrt}
  R_2(\a)\approx \text{const} - 3^{1/4}/\sqrt{\a_c-\a} .
\ee

The dependence of the average current as a function of the average superfluid velocity in the 2D case obtained from Eqs.\ \eqref{j-av-2D} and \eqref{R2} for $\aaa_*^\text{2D}/\aaa = 0.05$ is shown in Fig.~\ref{finalresults} by the red line.

\subsubsection{Region of strong fluctuations near criticality}

Our results for the average current, Eqs.\ \eqref{j-av-3D} and \eqref{j-av-2D}, are inapplicable in the vicinity of the bare critical current. Indeed, due to the power-law dependence of $R_d(\a)$ at $\a\to \a_c$ [Eqs.\ \eqref{R3sqrt} and \eqref{R2sqrt}], the average current has a maximum at $\a=\a_c-\delta \a_d$, where $\delta \a_d$ depends of the effective dimensionality:
\be
\label{dvd}
  \delta \a_3
  \sim 
  \left( \frac{\aaa_*^\text{3D}}{\aaa} \right)^{1/3} ,
\qquad
  \delta \a_2
  \sim 
  \left( \frac{\aaa_*^\text{2D}}{\aaa} \right)^{2/5} .
\ee
In a small region near the critical current, at $\a\gtrsim\a_c-\delta \a_d$, the derivative $d\corr{j}/d\a$ becomes negative, signaling instability of the corresponding branch. Physically it is related to the breakdown of the first-order perturbation theory at $\a_c-\a\lesssim \delta \a_d$. For such small deviations from $\a_c$, the inhomogeneity-induced self-energy for the fluctuation propagator $L$ becomes comparable to the lowest eigenvalue $\Lambda_-$ of the bare $L^{-1}$ that vanishes at $\a_c$ as discussed in Sec.\ \ref{SS:FP}. 

Hence, the region $\a_c-\a\lesssim\delta \a_d$ is a \emph{fluctuation region}, where system's properties are determined by summation of an infinite series of diagrams that will be analyzed elsewhere.
The first-order perturbative corrections \eqref{j-av-3D} and \eqref{j-av-2D} are valid outside the fluctuation region, at
\be
  \a_c-\a\gg\delta \a_d.
\ee
The fluctuation region is shaded in Fig.\ \ref{finalresults}.

Note that the singular behavior of $R_2(\a)$ in 2D given by Eq.~\eqref{R2sqrt} comes from large scales $r_*(\delta \a)\sim\xi/\sqrt{\delta \a}$. In order for the theory to be self-consistent, this scale should be smaller than the infra-red regularizing length $L_*\sim\xi(\aaa/\aaa_*^\text{2D})^{1/2}$ introduced in Sec.\ \ref{av-Delta2}. Taking $r_*$ at the border of the fluctuation region, $\delta\a\sim\delta\a_d$, we see that indeed $r_*(\delta \a_2)\ll L_*$ since we are always working in the limit $\aaa\gg \aaa_*^\text{2D}$.

\subsection{Supercurrent correlation function}

In order to characterize current fluctuations in the lowest order in inhomogeneity strength, it is sufficient to consider the first perturbative correction to the uniform current, $\delta\bj(\br)$. Substituting expansion \eqref{Delta-exp} into Eq.~\eqref{current}, we obtain
\be
  \delta \bj 
  = \jcoeff \Delta_0
  \left[ 
    \bA (\Delta_1 + \Delta_1^*) 
  + \frac{\nb \Delta_1 - \nb \Delta_1^*}{2i} 
  \right] .
\ee
With the help of Eq.\ \eqref{Delta1} we get for the $i$'th component of $\delta\bj(\bq)$:
\be
\label{djq}
  \frac{\delta j_i (\bq)}{j_0({\color{\revcolor}\a})}
  = 
  \frac{2 \aaa_1(\bq)}{\aaa} 
  \frac{q'^2\delta_{ix} - q'_i q'_x}{q'^4 + 2 (1-\a^2)q'^2 - 4\a^2 q_x'^2} ,
\ee
where $\bq'$ is the dimensionless momentum, see Eq.\ \eqref{q-dimless}.

Averaging over inhomogeneities with $r_c\ll\xi$, replacing $f_\tau(q)$ by $\fzeroq = f_\tau(q\,{=}\,0)$ and converting it to $\tau_*^{(d)}$ with the help of Eqs.\ \eqref{alpha*3D} and \eqref{alpha*2D}, we obtain the correlation function of current fluctuations for $d=2$, 3:
\be
\label{defC}
  \frac{\corr{\delta j_i (0) \delta j_j (\br)}}{j^2_0(a)}
  = 
  \biggl( \frac{\tau_*^{(d)}}{\tau} \biggr)^{2-d/2} C_{ij}^{(d)} (\br) ,
\ee
where the tensor $C_{ij}(\br)$ is defined as
\be
\label{C-gen}
  C_{ij}^{(d)}(\br) 
  = 
  \gamma_d \int 
  (d\bq') e^{i \bq' \br/\xi}
  \frac{(q'^2\delta_{ix} - q'_i q'_x)(q'^2\delta_{jx} - q'_j q'_x)}
  {[q'^4 + 2 (1-\a^2)q'^2 - 4\a^2 q_x'^2]^2} 
\ee
with $\gamma_d = 2^{6-d/2}\pi/d$.
The relative magnitude of current fluctuations \eqref{defC} is governed by the factor $(\tau_*^{(d)}/\tau)^{2-d/2}$, while the dimensionless tensor $C_{ij}^{(d)}(\br)$ describes their spatial and $v$ dependence. 
For a generic $\br$, all components of $C_{ij}^{(d)}(\br)$ are nonzero and typically cannot be obtained in a closed form. Below we present analytic results for the correlators at coincident points and their far asymptotics.

At \emph{coincident} points, the tensor $C_{ij}^{(d)}(0)$ is diagonal and uniaxial, with $x$ being the principal axis, and the momentum integration in Eq.\ \eqref{C-gen} can be done analytically. In the 2D geometry, 
\begin{gather}
\label{Cxx02}
  C_{xx}^\text{2D}(0) 
  = 
  \frac{1}{2 \color{\revcolor}\a^4}
  \left[(1-3\a^2)\sqrt{\frac{1-3\a^2}{1-\a^2}} - 1 + 4\a^2 \right] ,
\\
\label{Cyy02}
  C_{yy}^\text{2D}(0)
  =
  \frac{1}{2\left[ 1-2\a^2 + \sqrt{(1-\a^2)(1-3\a^2)}\right]} ,
\end{gather}
while in the 3D case
\begin{gather}
\label{Cxx03}
  \! C_{xx}^\text{3D}(0)
  =
  \frac{(3-22\a^2+51\a^4) \sigma(\a) -3(1-5\a^2)\rho(\a)}{2^{9/2}3 \a^5} ,
  \!
\\
\label{Cyy03}
  C_{yy}^\text{3D}(0)
  = \frac{(3-7\a^2)\rho(\a) - (3-14\a^2+11\a^4)\sigma(\a)}{2^{11/2}3 \a^5} ,
\end{gather}
where we introduced $\sigma(\a) = \arctan{\sqrt{2\a^2/(1-3\a^2)}}$ and $\rho(\a) = \sqrt{2\a^2(1-3\a^2)}$. By symmetry, $C_{zz}^\text{3D}(0) = C_{yy}^\text{3D}(0)$.
Equations \eqref{Cxx02}--\eqref{Cyy03} describe smooth monotonically increasing functions, taking finite positive values both at $\a=0$ and $\a=\a_c$.

\begin{figure}
\centerline{\includegraphics[width=0.9\columnwidth]{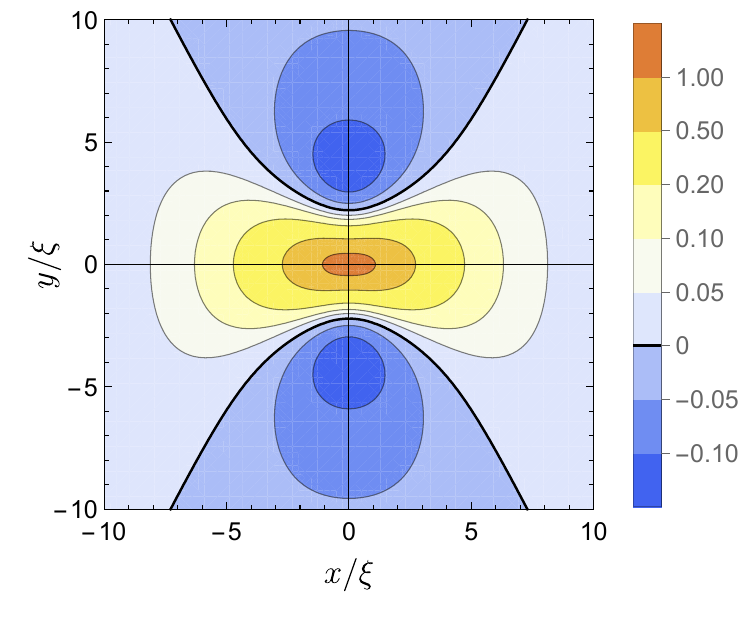}}
\caption{Function $C_{xx}^\text{2D}(\br)$ that determines the correlation function $\corr{\delta j_x (0) \delta j_x (\br)}$ of longitudinal currents in the 2D case at $\a/\a_c = 0.95$.}
\label{fluccurrent2x}
\end{figure}

The asymptotic behavior of the current correlation functions \emph{in the limit $r\to\infty$} can be obtained by neglecting the quartic term in the denominator of Eq.\ \eqref{cal-L(q)} and using the Laplace transform. Similar to Eq.\ \eqref{L-far}, we obtain a power-law decay with the exponent $d$. In the 2D case,
\begin{align}
  C^\text{2D}_{xx}(\br) 
  & = 
  \frac{\xi^2}{r^2} 
  \frac{(1-3\a^2)^{1/2} P^\text{2D}_x(\theta)}
    {(1-\a^2)^{3/2} \Pi^3(\theta)} ,
\\{}
  C^\text{2D}_{yy}(\br) 
  & = 
  \frac{\xi^2}{r^2} 
  \frac{P^\text{2D}_y(\theta)}
    {(1-\a^2)^{1/2} (1-3\a^2)^{1/2} \Pi^3(\theta)} ,
\end{align}
where $\Pi(\theta)$ and $P_{x,y}(\theta)$ are given by
\be
  \Pi(\theta) = 1-3\a^2 + 2\a^2 \cos^2\theta
\ee
and
\begin{align}
  P^\text{2D}_x(\theta)
  = &
  -(1-3\a^2)^2-4(1-3\a^2)\cos^2\theta
\nonumber \\{}
  & + 4(2-6\a^2+3\a^4) \cos^4\theta ,
\\
  P^\text{2D}_y(\theta) = & -(1-3\a^2)^2+4(2-9\a^2+9\a^4) \cos^2\theta 
\nonumber \\{}
  & -4(2-8\a^2+7\a^4) \cos^4\theta .
\end{align}
In the 3D case, the longitudinal and transverse correlators have the asymptotic behavior
\begin{align}
  C^\text{3D}_{xx}(\br) & = 
  \frac{\xi^3}{r^3} 
  \frac{(1-3\a^2)P^\text{3D}_x(\theta)}{\sqrt{18}(1-\a^2)^{3/2} \Pi^{7/2}(\theta)} ,
\\{}
  C^\text{3D}_{yy}(\br) + C^\text{3D}_{zz}(\br) & = 
  \frac{\xi^3}{r^3} 
  \frac{P^\text{3D}_\perp(\theta)}{\sqrt{18}(1-\a^2)^{1/2} \Pi^{7/2}(\theta)} ,
\end{align}
with
\begin{align}
  P^\text{3D}_x(\theta)
  = &
  -(1-3\a^2)^2-2(3-10\a^2+3\a^4)\cos^2\theta 
\nonumber \\{}
  & +(15-42\a^2+23\a^4)\cos^4 \theta ,
\\
  P^\text{3D}_\perp(\theta) 
  = & -(1-3\a^2)^2+4(3-13\a^2+12\a^4)\cos^2\theta 
\nonumber \\{}
  & -(15-54\a^2+43\a^4)\cos^4 \theta .
\end{align}

\begin{figure}
\centerline{\includegraphics[width=0.9\columnwidth]{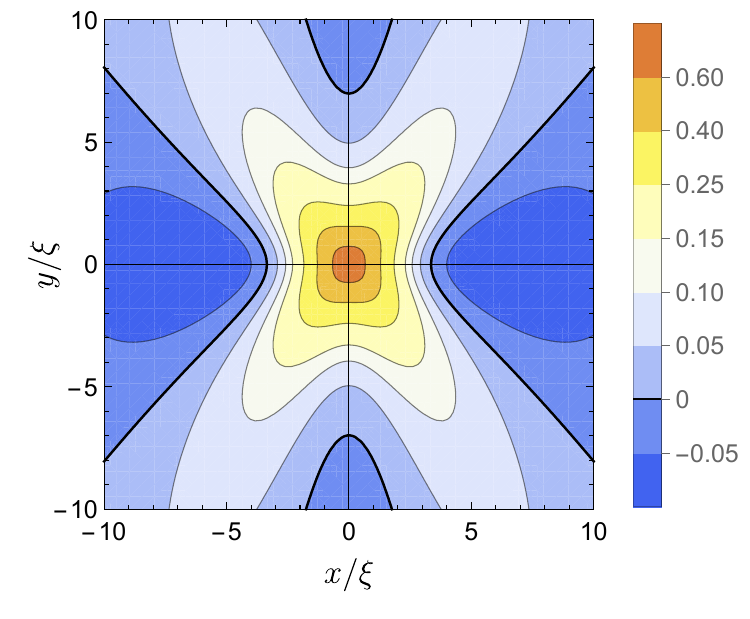}}
\caption{Function $C_{yy}^\text{2D}(\br)$ that determines the correlation function $\corr{\delta j_y (0) \delta j_y (\br)}$ of transverse currents in the 2D case at $\a/\a_c = 0.95$.}
\label{fluccurrent2y}
\end{figure}

In the 2D geometry, the spatial dependence of the longitudinal, $\corr{\delta j_x (0) \delta j_x (\br)}$, and transverse, $\corr{\delta j_y (0) \delta j_y (\br)}$, current correlation functions at $\a/\a_c = 0.95$ is shown in Figs.\ \ref{fluccurrent2x} and \ref{fluccurrent2y}, respectively. Both functions are notably anisotropic and demonstrate anticorrelations at large $y$. Also negative is $\corr{\delta j_y (0) \delta j_y (\br)}$ for large $x$, in accordance with Eqs.\ \eqref{Cxx02} and \eqref{Cyy02}.

\subsection{Numerical simulation}

Figure \ref{F:modelcurrent} demonstrates the result of numerical simulation of the modulus of the order parameter (color map) and supercurrent (current flow lines) in the 2D geometry at $\tau_*^\text{2D}/\tau=0.01$.
\color{\revcolor}
To generate these images we considered a discrete version of the model on a grid of size $128\times128$ with periodic boundary conditions. The values of $\tau_1(\br_i)$ at each site were generated randomly and independently from the Gaussian distribution. The coherence length $\xi$ was assumed to be 10 lattice spacings, hence $\xi/r_c=10$. The first corrections to the order parameter and supercurrent in the momentum representation were obtained from Eqs.\ \eqref{Delta1} and \eqref{djq}, where $\tau_1(\bq_i)$ was calculated as the fast Fourier transform of $\tau_1(\br_i)$. Making then the inverse Fourier transform, we obtain $|\Delta(\br_i)|$ and $\bj(\br_i)$ shown in Fig.\ \ref{F:modelcurrent}.

\color{black}
The left and right panels are obtained for the same realization of $\tau_1(\br)$ but at different values of the average superfluid velocity: $\a/\a_c \to 0$ and $\a/\a_c=0.85$, respectively.
One can clearly see that increasing the current enhances fluctuations of $|\Delta(\mathbf{r})|$ and renders the supercurrent pattern more inhomogeneous. The latter effect can be quantified by the root mean square of the angle $\vartheta$ between the local supercurrent and the direction of the average superfluid velocity. Since $\vartheta=j_y/j_0$, $\corr{\vartheta^2}$ is proportional to $C_{yy}(0)$ given by Eq.~\eqref{Cyy02}. 
This correlation function at $v=0.85 \, v_c$ is 2.05 times larger than at $v=0$. Hence the typical angle $\corr{\vartheta^2}^{1/2}$ at $v=0.85 \, v_c$ should be 43\% larger than that at a vanishing bias, in accordance with Fig.\ \ref{F:modelcurrent}.

\color{\revcolor}
\section{Discussion and Conclusion}
\label{S:concl}

\color{black}
In this paper we have developed an analytical approach to the problem of the supercurrent flow in inhomogeneous superconductors. Our theory based on the Ginzburg-Landau expansion is applicable in the vicinity of $T_c$, where any microscopic mechanism of inhomogeneity formation manifests itself as a random correction of the form $\tau_1(\br)|\Delta(\br)|^2$ to the quadratic term in the free energy. The resulting random-temperature $\phi^4$ theory with the complex order parameter is analyzed perturbatively.

The key element of the theory is the fluctuation propagator, which describes the response of the amplitude and phase modes of the order parameter. In the presence of a superflow, these modes get coupled already at the Gaussian level. As a consequence of such a hybridization, even the amplitude mode becomes long-range, inheriting a power-law decay of the Goldstone phase mode. As a result, all correlation functions demonstrate a power-law rather than exponential attenuation at large distances.

In the most relevant case of short-range in\-ho\-mo\-ge\-ne\-ities ($r_c\ll\xi$), the theory contains a loop diagram divergent in the ultraviolet. We argue that this divergency should be absorbed into the redefinition of the critical temperature, similar to the shift of $T_c$ due to thermal fluctuations. After proper regularization, one arrives at a theory well defined at small scales, which depends only on the zero Fourier component $\fzeroq = f_\tau(q\,{=}\,0)$ of the $\tau_1$ correlation function. Within this theory we calculate the dependence of the correction to the average supercurrent $\corr{\bj}$ and its correlation function $\corr{j_ij_j}$ on the \emph{average} superfluid velocity $v=\xi\corr{\tilde{\mathbf{v}}_s}$, the quantity that can be accessed experimentally by applying a phase difference across the sample. At a qualitative level, the obtained expressions demonstrate suppression of the average superfluid current by inhomogeneity and growth of spatial fluctuations of $|\Delta(\br)|$ and $\mathbf{j}(\br)$ with increasing the phase gradient imposed.

The perturbation theory developed is applicable at $\tau\gg\tau_*$, where $\tau$ is the dimensionless distance from $T_c$, and $\tau_*$ is the effective inhomogeneity strength given by Eq.\ \eqref{alpha*d}. Another limitation is imposed by the value of the supercurrent, which should not be very close to the critical current, $\a_c-\a\gg\delta \a_d$, where the width of the fluctuation region is given by Eq.~\eqref{dvd}. Finally, thermal fluctuations will not smear frozen-in-space patterns of $|\Delta(\br)|$ and $\textbf{j}(\br)$ provided $\tau\gg\text{Gi}$.

\color{\revcolor}
One of the experimentally observable consequences of the developed theory is Eq.\ \eqref{dns2D}, which describes the suppression of the superfluid density by inhomogeneities. 

An intriguing proposal is trying to detect the magnetic field generated by a non-uniform supercurrent flow using the SQUID-on-tip technique developed in Ref.\ \cite{Zeldov-SOT} for precise vortex imaging. In a uniform situation, the current $\bj$ flowing along the $x$ axis generates a normal magnetic field $B_z$ changing in the $y$ direction.
In a non-uniform situation, the inhomogeneous part of the current $\delta\bj(\br)$ will produce a random correction $\delta B_z(x,y,z)$ to the normal field that will depend on $x$ as well. Solving Maxwell's equation $\rot \delta\mathbf{B}=(4\pi/c) \delta\bj(x,y) d_z \delta(z)$, taking $\delta\bj$ from Eq.\ \eqref{djq} and averaging over disorder, we obtain for the root mean square of $\delta B_z$ near the surface:
\be
  \sqrt{\corr{\delta B_z^2}}
  = 
  \frac{2\pi^3eg\tau T_c}{7\zeta(3)c\xi_0}
  (1-v^2)v
  \sqrt{\tau_*
  [C_{xx}(0)+C_{yy}(0)]} ,
\ee
where $g=h/e^2R_\Box$ is the dimensionless film resistance, and the local correlations functions $C_{ii}(0)$ are given by Eqs.\ \eqref{Cxx02} and \eqref{Cyy02}. The largest field is achieved near the critical current ($\a\to1/\sqrt{3}$), where we get
\be
  \mathop{\rm rms}B_z\text{[G]}
  \approx
  \frac{4 \tau \sqrt{\tau_*} \, T_c\text{[K]}}
  {R_\Box\text{[k$\Omega$]} \, \xi_0\text{[nm]}} .
\ee
For a 3-nm-thick NbN film with $R_\Box\approx3$ k$\Omega$, $T_c\approx6$ K, $\xi_0\approx6$ nm \cite{Astafiev} and assumed inhomogeneity strength $\tau_*\approx 0.1$, extrapolation to low temperatures ($\tau\to1$) yields $\mathop{\rm rms}B_z\approx 0.4$ G. Thus the resulting local fluctuations of the normal magnetic field are comparable to Earth's magnetic field.

Whether such a quenched magnetic field $\delta B_z(x,y)$ can be detected by a SQUID-on-tip magnetometer depends on the size of the SQUID loop $L_\text{loop}$, which was about 100 nm in the experiment \cite{Zeldov-SOT}. Since $L_\text{loop}\gg\xi_0$, the magnetic flux through the loop will be a sum of independent contributions of many puddles. At small currents their number can be estimated as $N\sim(L_\text{loop}/\xi_0)^2$, while near the critical current it reduces to $N\sim L_\text{loop}/\xi_0$ due to the long-range correlations. Hence the flux $\Phi$ through the loop will fluctuate with $\mathop{\rm rms}\Phi = L_\text{loop}^2 \mathop{\rm rms}B_z/\sqrt{N} \approx 5\times10^{-5}\,\Phi_0$, where $\Phi_0$ is the superconducting flux quantum. This value is close to the detection threshold for SQUID devices, making detection of the magnetic field generated by an inhomogeneous superflow a plausible, but challenging task.

\color{black}
A natural development of our theory would be to generalize it to the case of arbitrary temperatures. Then instead of using the GL expansion, one has to consider the Usadel equation for the quasiparticle Green function and the self-consistency equation for the order parameter. Finally, a challenging and experimentally relevant in an SNSPD context problem is to understand the formation of thermal phase slips in inhomogeneous superconductors that will be considered elsewhere.

\color{black}

\begin{acknowledgments}

We are grateful to V. V. Lebedev for discussing the critical temperature renormalization by thermal fluctuations.
The work of M.A.S. and O.B.Z. was supported by the Russian Science
Foundation under Grant No.\ 23-12-00297 and by the Foundation for the Advancement of Theoretical Physics and Mathematics ``BASIS''.

\end{acknowledgments}

\appendix

\section{Amplitude response in 2D}
\label{A:L2D}

Here we derive the intermediate asymptotics \eqref{L2Dint} of the amplitude propagator ${\cal L}(\br)$ in the 2D geometry at $\a\to \a_c$. The starting point is Eq.\ \eqref{cal-L(q)} for ${\cal L}(\bq)$, where it is convenient to integrate first over $q'_y$. As a function of $q'_y$, ${\cal L}(\bq)$ has four poles located at $ik_\pm$ and $-ik_\pm$, where $k_\pm = \sqrt{1-\a^2+q'^2_x \pm \rho}$ and $\rho=\sqrt{(1-\a^2)^2 + 4\a^2 q'^2_x}$. Integrating over $q'_y$, we get an expression valid for all $v$:
\be
  {\cal L}_\text{2D}(\br) 
  = 
  \frac{1}{8\pi \aaa\xi^2} 
  \int_{-\infty}^{\infty} d q'_x
  (I_++I_-) e^{i q'_x x/\xi} ,
\ee
where
\be
I_\pm
=
  \pm \frac{1-v^2 \pm \rho}{\rho k_\pm} e^{-k_\pm |y|/\xi} 
.
\ee
The analytic structure of $I_+$ and $I_-$ is different. At $v\to v_c$ (more precisely, at $v>1/\sqrt5$), both terms have branch cuts $(-i\infty,-i\kappa_2)$ and $(i\kappa_2,i\infty)$, while the term $I_-$ has also a branch cut $(-i\kappa_1,i\kappa_1)$, which crosses the real axes. The position of the branching points is determined by $\kappa_1=\sqrt{2(1-3v^2)}$ and $\kappa_2=(1-v^2)/2v$, with $\kappa_1<\kappa_2$ and $\kappa_1$ vanishing at $v_c$. Note also that $k_-$ vanishes at $q'_x=0$, while $k_+$ remains finite. By deforming the integration contour we conclude that the asymptotic behavior of ${\cal L}_\text{2D}(\br)$ at $r\gg\xi$ is determined by $I_-$, where it suffices to keep only the contribution of the contour $C$ encompassing the cut $(0,i\kappa_1)$:
\be
\label{L2D-contour}
  {\cal L}_\text{2D}(\br) 
  \approx
  - 
  \frac{1}{8\pi \aaa\xi^2} 
  \oint_{C} d q'_x \, 
  \frac{1-\a^2 - \rho}{\rho k_-} e^{i q'_x x/\xi - k_- |y|/\xi} .
\ee
In the leading order in $v_c-v$ and $q'_x$, we have 
$\rho=2/3$, 
$\kappa_1^2=4\sqrt3(v_c-v)$,
\begin{gather}
  1-v^2-\rho
  =
  - q_x'^2 ,
\\
  k_-^2 = 
  (3/4)
  q_x'^2 (\kappa_1^2+q_x'^2) .
\end{gather}

Now introducing $q'_x=i\kappa_1t\pm0$, we write the contour integral \eqref{L2D-contour} as (here we assume $x,y>0$)
\be
\label{L2D-A}
  {\cal L}_\text{2D}(\br) 
  =
  - 
  \frac{\sqrt{3}\kappa_1}{4\pi \aaa\xi^2} 
  \re
  \int_0^1 
  \frac{dt \, t \, e^{-\kappa_1t (x + i \sqrt3 \kappa_1 \sqrt{1-t^2} y/2)/\xi}}{\sqrt{1-t^2}} 
  .
\ee
At largest $r$, the main contribution comes from $t\ll1$, and we reproduce the power-law far asymptotics ${\cal L}^\text{far}(\br)$ given by Eq.~\eqref{L-far}. As $r$ is decreased, the relevant values of $t$ increase and become $t\sim1$ when $r$ reaches $r_c(\theta)$ [Eq.~\eqref{condition}], which is still much larger than $\xi$. At $r\lesssim r_c(\theta)$, the exponent in Eq.\ \eqref{L2D-A} can be discarded, and we arrive at the $\br$-independent intermediate asymptotics \eqref{L2Dint}.

\end{document}